\documentclass[aps,preprint,nofootinbib]{revtex4}
\usepackage{amsmath}
\usepackage{graphicx}
\usepackage{color}
\usepackage{slashed}
\usepackage[aboveskip=-5pt]{subcaption}
\usepackage{booktabs}
\usepackage{mathtools}
\usepackage{multirow}
\usepackage{makecell}
\allowdisplaybreaks

\begin{document}

\title{NLO QCD corrections to the $B_c$-pair hadroproduction}

\author{\vspace{1cm} Zi-Qiang Chen$^1$\footnote[1]{chenziqiang13@gzhu.edu.cn}, Long-Bin Chen$^1$\footnote[2]{chenlb@gzhu.edu.cn, corresponding author} and Cong-Feng Qiao$^{2}$\footnote[3]{qiaocf@ucas.ac.cn, corresponding author}}

\affiliation{$^1$School of Physics and Materials Science, Guangzhou University, Guangzhou 510006, China \\
$^2$School of Physical Sciences, University of Chinese Academy of Science, Yuquan Road 19A, Beijing 100049, China}

\begin{abstract}
\vspace{0.5cm}

The $B_c$ meson pair, including pairs of pseudoscalar states and vector states, productions in proton-proton collisions are investigated at the next-to-leading order (NLO) accuracy in the nonrelativistic quantum chromodynamics factorization formalism.
The corresponding cross sections at the Large Hadron Collider (LHC) with $\sqrt{s}=14\; \text{TeV}$ are evaluated.
Numerical results indicate that the NLO corrections are substantial, and even dominate over the leading order contributions.
Considering the predicted cross sections are sizable, the $B_c$-pair production is expected to be observable at the High-Luminosity LHC experiment.

\end{abstract}
\maketitle

\newpage

\section{Introduction}

As a multiscale system consisting of two heavy quarks,
heavy quarkonium presents an ideal laboratory for testing the interplay between perturbative and nonperturbative quantum chromodynamics (QCD) within a controlled environment.
Due to the large mass of heavy quark, quarkonium production process is assumed to be factorisable into two steps:
First, a heavy quark-antiquark pair whose invariant mass near the bound state mass is produced perturbatively, then the pair hadronizes into a quarkonium state nonperturbatively.
The nonrelativistic QCD (NRQCD) factorization formalism, which proposed by Bodwin, Braaten, and Lepage \cite{Bodwin:1994jh},
provides a solid foundation for the theoretical study of quarkonium production.
In NRQCD factorization approach, the nonperturbative hadronization effects are encoded into the long distance matrix elements (LDMEs), which are sorted in powers of the relative velocity $v$ of the heavy quarks in the bound state.
Although the quarkonium production has been extensively investigated at various colliders, the existing researches are still not sufficient to pin down the color-octet (CO) LDMEs (see, for example Refs. \cite{Andronic:2015wma,Lansberg:2019adr} for a review).

Over the past two decades, there has been increasing interest in quarkonium-pair hadroproduction \cite{LHCb:2011kri,CMS:2014cmt,CMS:2016liw,ATLAS:2016ydt,LHCb:2016wuo,CMS:2020qwa,
ALICE:2023lsn,LHCb:2023ybt,LHCb:2023wsl,Baranov:1997wy,Qiao:2009kg,Li:2009ug,Ko:2010xy,Kom:2011bd,
Berezhnoy:2012tu,Martynenko:2012tf,Li:2013csa,Sun:2014gca,Trunin:2015uma,Baranov:2015cle,Likhoded:2016zmk,
He:2015qya,Lansberg:2017dzg,He:2019qqr,Scarpa:2019fol,Lansberg:2020rft,Prokhorov:2020owf,Baranov:2022zgg,Sun:2023exa} initiated in Ref.\cite{Qiao:2002}, as it provides a sensitive test for NRQCD factorization formalism.
Due to the high reconstruction efficiency of $J/\psi$ meson, most researches focus on $J/\psi$-pair production,
for which, the contributions of double $c\bar{c}({}^3S_1^{[1]})$ and double $c\bar{c}({}^1S_0^{[8]})$ intermediate states are calculated up to next-to-leading order (NLO) QCD accuracy \cite{Sun:2014gca,Sun:2023exa}.
Nevertheless, there are many challenges in understanding the underlying production mechanism.
First, the full NRQCD prediction requires to include the contributions of all possible pairing of $c\bar{c}({}^1S_0^{[8]})$, $c\bar{c}({}^3S_1^{[1,8]})$, and $c\bar{c}({}^3P_J^{[1,8]})$ intermediate states;
while the higher order calculations to these processes are hard to proceed, especially for double $c\bar{c}({}^3P_J^{[1,8]})$ channel \cite{He:2018hwb}.
Second, the results of Refs. \cite{Sun:2014gca,Sun:2023exa} indicate that the NLO corrections may dominant over the leading order (LO) contributions, which exhibits poor perturbative convergence --- one may worry that the calculation at NLO is not accurate enough to make a reliable prediction.
Third, the $J/\psi$-pair hadroproduction may serve as a probe to double parton scattering (DPS) mechanism \cite{Kom:2011bd};
while current researches are inadequate to draw any concrete conclusion, since the predictions to the single parton scattering (SPS) contribution are fraught with uncertainties as well.

As the Large Hadron Collider (LHC) may transit to its high luminosity operation in 2029 and beyond,
more interesting processes, like the $B_c$-pair hadroproduction, can be measured with high accuracy.
Compared with $J/\psi$-pair hadroproduction, the production mechanism of $B_c$-pair hadroproduction is much simpler.
Both the DPS and CO contributions are expected to be small in comparison with the traditional color-singlet SPS contribution, due to the fact that the $B_c$ meson needs to be produced in accompany with an additional $b\bar{c}$ pair.
Hence the $B_c$-pair hadroproduction provides a complementary approach to clarifying the puzzles in $J/\psi$-pair hadroproduction, and more clear conclusions can be expected due to its simple production mechanism.
In Refs. \cite{Baranov:1997wy,Li:2009ug}, the LO calculation of $B_c$-pair hadroproduction is performed in the NRQCD factorization framework,
and in Ref. \cite{Trunin:2015uma}, the relativistic relativistic correction is carried out by using the relativistic quark model approach.
Considering the fact that the NLO QCD corrections to quarkonium-pair production processes are normally significant \cite{Sun:2014gca,Sun:2023exa},
in this work, we calculate the NLO QCD corrections to $B_c$-pair hadroproduction.
Various of $S$-wave $B_c$ states, including configurations of $B_c^++B_c^-$, $B_c^++B_c^{*-}$ \footnote{The productions of $B_c^++B_c^{*-}$ and $B_c^{*+}+B_c^-$ are related by a charge conjugation transformation. Their cross sections are exactly the same.}, and $B_c^{*+}+B_c^{*-}$, are taken into account.
Note, here after for simplicity, $B_c$ represents for both pseudoscalar $B_c$ and vector $B_c^*$,
the latter may overwhelmingly decay to the pseudoscalar state, unless specifically mentioned.

The rest of the paper is organized as follows.
In Sec. II, we present the LO calculation of $B_c$-pair production in proton-proton collisions.
In Sec. III, some technical details in the calculation of NLO corrections are given.
In Sec. IV, the numerical evaluation for concerned processes is performed at NLO QCD accuracy.
The last section is remained for summary and conclusions.

\section{The LO cross sections}
\begin{figure}
\begin{subfigure}{\textwidth}
\includegraphics[scale=0.5]{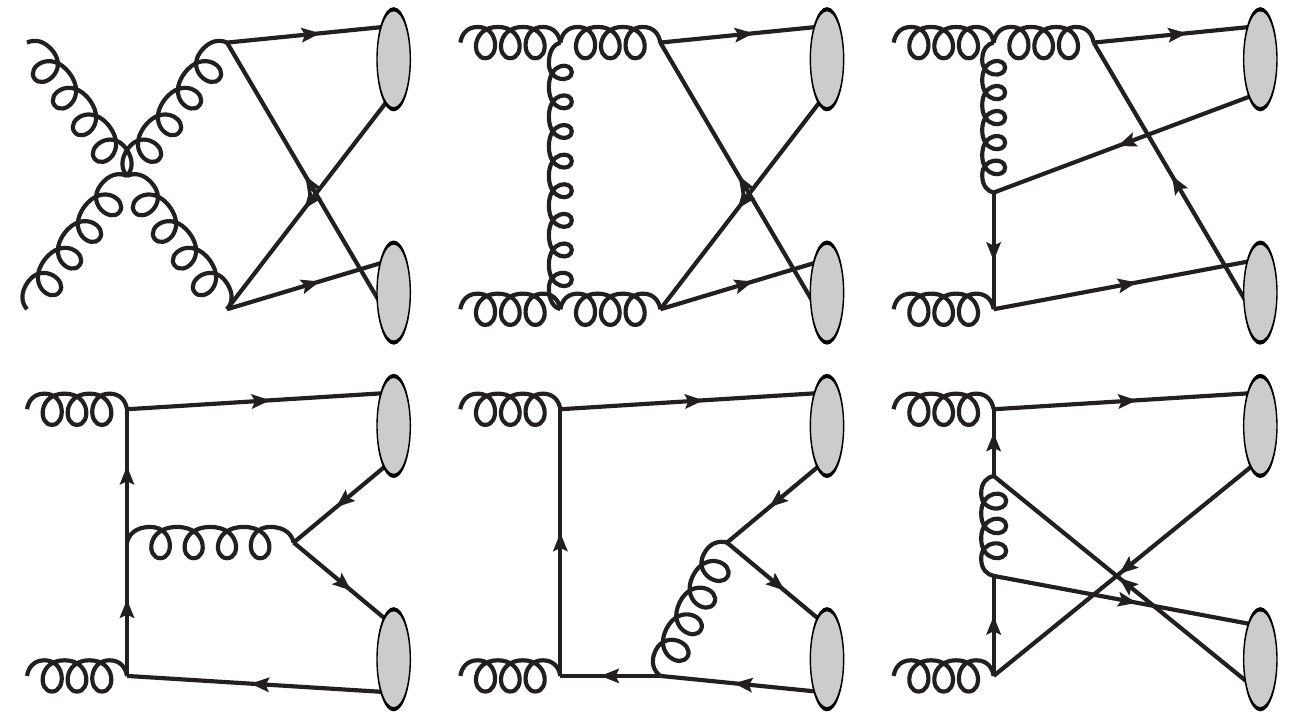}
\caption{}
\end{subfigure}
\begin{subfigure}{\textwidth}
\includegraphics[scale=0.5]{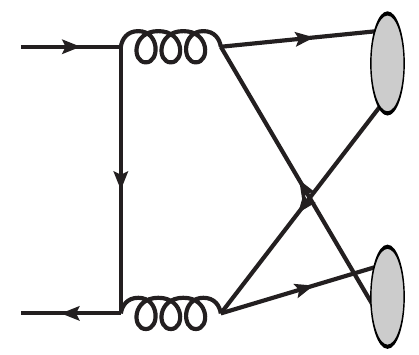}
\caption{}
\end{subfigure}
\caption{The typical tree-level Feynman diagrams for (a) $gg$-induced process, (b) $q\bar{q}$-induced process.
The remaining diagrams can be obtained by reversing the quark lines, or interchanging the initial gluons.}
\label{fig_FeynTree}
\end{figure}

At the LO in $\alpha_s$, $B_c$-pair hadroproduction receives contributions from two partonic processes, namely
the $gg$-induced process
\begin{equation}
g(p_1)+g(p_2)\to B_c^+(k_1)+B_c^-(k_2), \label{eq_LOprocess1}
\end{equation}
and the $q\bar{q}$-induced process
\begin{equation}
q(p_1)+\bar{q}(p_2)\to B_c^+(k_1)+B_c^-(k_2),\label{eq_LOprocess2}
\end{equation}
whose Feynman Diagrams are shown in Fig. \ref{fig_FeynTree}.
Since the $q\bar{q}$-induced process is suppressed by quark parton distribution functions (PDFs), it was missed by previous LO studies \cite{Baranov:1997wy,Li:2009ug,Trunin:2015uma}.
While at NLO, to eliminate the collinear singularities of the real emission process $g+q(\bar{q})\to B_c^++B_c^-+q(\bar{q})$, the $q\bar{q}$-induced process should be convoluted with the scale dependent PDFs.
Hence, for a thorough study, both $gg$- and $q\bar{q}$-induced processes should be taken into account.
In fact, our numerical results show that at LO the contribution of $q\bar{q}$-induced process may reach $10\%$ in some cases, which is not really negligible.

According to the NRQCD factorization formalism, the LO partonic cross section can be formulated as
\begin{equation}
d\hat{\sigma}_{\rm LO}^{i+j\to B_c^++B_c^-}=\frac{|\Psi(0)|^4}{2s}\overline{\sum}\left|\mathcal{M}^{i+j\to [b\bar{c}]+[\bar{b}c]}_{\rm tree}\right|^2d{\rm PS}_2,
\end{equation}
where $\Psi(0)$ is the wave function of $B_c$ meson at the origin,
$s=(p_1+p_2)^2$ is the center-of-mass energy squared of the colliding partons $i$ and $j$,
$\overline{\sum}$ sums (averages) over the polarizations and colors of the final (initial) state particles,
$\mathcal{M}^{i+j\to [b\bar{c}]+[\bar{b}c]}_{\rm tree}$ is the corresponding LO partonic amplitude,
and $d{\rm PS}_2$ stands for the 2-body phase space.

The partonic amplitude can be computed by using the covariant projection operator method.
At the leading order of the relative velocity expansion, it is legitimate to take $m_{B_c}=m_b+m_c$, $p_{B_c}=p_c+p_b=(1+\frac{m_c}{m_b})p_b$, where $m_{B_c}$ and $p_{B_c}$ denote the mass and momentum of the $B_c$ meson respectively.
The spin and color projection operators take the form
\begin{equation}
\Pi(n)=\frac{1}{2\sqrt{m_{B_c}}}\varepsilon(n)(\slashed p_{B_c}+m_{B_c})\otimes \frac{1_c}{\sqrt{N_c}},
\end{equation}
where $\varepsilon(^1S_0)=\gamma_5$, $\epsilon(^3S_1)=\slashed\epsilon$, and $\epsilon$ represents the polarization vector of $B_c^*$ meson.
The $1_c$ stands for the unit color matrix, and $N_c=3$ is the number of colors in QCD.

The LO calculation is straightforward, and the analytic expressions for $\overline{\sum}|\mathcal{M}|^2$ are pretty simple.
By introducing the dimensionless variables $\hat{s}=(p_1+p_2)^2/m_{B_c}^2$, $\hat{v}=-(p_1-p_2)\cdot (k_1-k_2)/m_{B_c}^2$, $\hat{w}=\sqrt{\hat{s}^2-\hat{v}^2}$, and $r=m_b/(m_b+m_c)$, we have
\begin{align}
\label{eq_ggPP}
&\overline{\sum}\left|\mathcal{M}^{g+g\to {}^1S_0+{}^1S_0}_{\rm tree}\right|^2=\tfrac{g_s^8}{m_{B_c}^6 r^4(1-r)^4}\big[ \tfrac{157}{36 \hat{s}^4}-\tfrac{374}{9 \hat{s}^3 \hat{w}^2}+\tfrac{22}{27 \hat{s}^3}+\tfrac{32 \hat{s}^2}{81 \hat{w}^8}+\tfrac{968}{9 \hat{s}^2 \hat{w}^4}-\tfrac{241}{27 \hat{s}^2 \hat{w}^2}+\tfrac{4}{81 \hat{s}^2}\nonumber  \\
&-\tfrac{16 \hat{s}}{9 \hat{w}^6}+\tfrac{860}{27 \hat{s} \hat{w}^4}-\tfrac{4}{9 \hat{s} \hat{w}^2}-\tfrac{352}{27 \hat{w}^6}+\tfrac{178}{81 \hat{w}^4}+\tfrac{1}{r(1-r)}\big(\tfrac{11 \hat{w}^2}{24 \hat{s}^5}+\tfrac{\hat{w}^2}{18 \hat{s}^4}-\tfrac{170}{27 \hat{s}^4}+\tfrac{844}{27 \hat{s}^3 \hat{w}^2}-\tfrac{731}{648 \hat{s}^3}\nonumber \\
&-\tfrac{1408}{27 \hat{s}^2 \hat{w}^4}+\tfrac{613}{81 \hat{s}^2 \hat{w}^2}-\tfrac{1}{18 \hat{s}^2}+\tfrac{40 \hat{s}}{81 \hat{w}^6}-\tfrac{1340}{81 \hat{s} \hat{w}^4}+\tfrac{17}{36 \hat{s} \hat{w}^2}+\tfrac{256}{81 \hat{w}^6}-\tfrac{11}{9 \hat{w}^4}\big)+\tfrac{1}{r^2(1-r)^2}\big(\tfrac{\hat{w}^4}{64 \hat{s}^6}-\tfrac{17 \hat{w}^2}{72 \hat{s}^5}\nonumber \\
&-\tfrac{\hat{w}^2}{32 \hat{s}^4}+\tfrac{257}{162 \hat{s}^4}-\tfrac{416}{81 \hat{s}^3 \hat{w}^2}+\tfrac{13}{36 \hat{s}^3}+\tfrac{512}{81 \hat{s}^2 \hat{w}^4}-\tfrac{13}{9 \hat{s}^2 \hat{w}^2}+\tfrac{1}{64 \hat{s}^2}+\tfrac{160}{81 \hat{s} \hat{w}^4}-\tfrac{1}{8 \hat{s} \hat{w}^2}+\tfrac{41}{162 \hat{w}^4}\big)\big],\\
\label{eq_ggPV}
&\overline{\sum}\left|\mathcal{M}^{g+g\to {}^1S_0+{}^3S_1}_{\rm tree}\right|^2=\tfrac{g_s^8 (1-2 r)^2}{m_{B_c}^6 r^6(1-r)^6}\big[-\tfrac{\hat{w}^2}{64 \hat{s}^4}+\tfrac{31}{648 \hat{s}^3}+\tfrac{1}{18 \hat{s}^2 \hat{w}^2}+\tfrac{1}{64 \hat{s}^2}+\tfrac{1}{648 \hat{s} \hat{w}^2}-\tfrac{41}{162 \hat{w}^4}\big],\\
\label{eq_ggVV}
&\overline{\sum}\left|\mathcal{M}^{g+g\to {}^3S_1+{}^3S_1}_{\rm tree}\right|^2=\tfrac{g_s^8}{m_{B_c}^6 r^4(1-r)^4}\big[\tfrac{32 \hat{s}^4}{81 \hat{w}^8}+\tfrac{157}{12 \hat{s}^4}-\tfrac{374}{3 \hat{s}^3 \hat{w}^2}-\tfrac{143}{81 \hat{s}^3}+\tfrac{32 \hat{s}^2}{27 \hat{w}^8}-\tfrac{32 \hat{s}^2}{9 \hat{w}^6}+\tfrac{968}{3 \hat{s}^2 \hat{w}^4}\nonumber \\
&+\tfrac{143}{27 \hat{s}^2 \hat{w}^2}+\tfrac{4}{81 \hat{s}^2}+\tfrac{80 \hat{s}}{81 \hat{w}^6}-\tfrac{236}{27 \hat{s} \hat{w}^4}+\tfrac{109}{81 \hat{s} \hat{w}^2}-\tfrac{352}{9 \hat{w}^6}+\tfrac{92}{27 \hat{w}^4}
+\tfrac{1}{r(1-r)}\big(\tfrac{11 \hat{w}^2}{8 \hat{s}^5}+\tfrac{\hat{w}^2}{18 \hat{s}^4}-\tfrac{170}{9 \hat{s}^4}+\tfrac{844}{9 \hat{s}^3 \hat{w}^2}\nonumber \\
&-\tfrac{475}{216 \hat{s}^3}+\tfrac{64 \hat{s}^2}{81 \hat{w}^6}-\tfrac{1408}{9 \hat{s}^2 \hat{w}^4}+\tfrac{1021}{81 \hat{s}^2 \hat{w}^2}-\tfrac{1}{18 \hat{s}^2}+\tfrac{40 \hat{s}}{27 \hat{w}^6}-\tfrac{2 \hat{s}}{9 \hat{w}^4}-\tfrac{1972}{81 \hat{s} \hat{w}^4}+\tfrac{275}{324 \hat{s} \hat{w}^2}+\tfrac{256}{27 \hat{w}^6}-\tfrac{61}{27 \hat{w}^4}\big)\nonumber \\
&+\tfrac{1}{r^2(1-r)^2}\big(\tfrac{3 \hat{w}^4}{64 \hat{s}^6}-\tfrac{17 \hat{w}^2}{24 \hat{s}^5}-\tfrac{\hat{w}^2}{8 \hat{s}^4}+\tfrac{257}{54 \hat{s}^4}-\tfrac{416}{27 \hat{s}^3 \hat{w}^2}+\tfrac{191}{162 \hat{s}^3}+\tfrac{512}{27 \hat{s}^2 \hat{w}^4}-\tfrac{38}{9 \hat{s}^2 \hat{w}^2}+\tfrac{5}{64 \hat{s}^2}+\tfrac{160}{27 \hat{s} \hat{w}^4}\nonumber \\
&-\tfrac{241}{648 \hat{s} \hat{w}^2}+\tfrac{41}{162 \hat{w}^4}\big)\big],
\end{align}
for $gg$-induced process, and
\begin{align}
&\overline{\sum}\left|\mathcal{M}^{q+\bar{q}\to {}^1S_0+{}^1S_0}_{\rm tree}\right|^2=\tfrac{g_s^8 }{m_{B_c}^6 r^6(1-r)^6 }\big[-\tfrac{2 \hat{v}^4}{243 \hat{s}^6}-\tfrac{8 \hat{v}^2}{243 \hat{s}^5}+\tfrac{2 \hat{v}^2}{243 \hat{s}^4} \big], \\
\label{eq_qqPV}
&\overline{\sum}\left|\mathcal{M}^{q+\bar{q}\to {}^1S_0+{}^3S_1}_{\rm tree}\right|^2=\tfrac{g_s^8 (1-2 r)^2}{m_{B_c}^6 r^6(1-r)^6}\big[-\tfrac{2 \hat{v}^2}{243 \hat{s}^4}+\tfrac{8}{243 \hat{s}^3}+\tfrac{2}{243 \hat{s}^2}\big], \\
\label{eq_qqVV}
&\overline{\sum}\left|\mathcal{M}^{q+\bar{q}\to {}^3S_1+{}^3S_1}_{\rm tree}\right|^2=\tfrac{g_s^8}{m_{B_c}^6 r^6(1-r)^6}\big[-\tfrac{2 \hat{v}^4}{81 \hat{s}^6}-\tfrac{8 \hat{v}^2}{81 \hat{s}^5}+\tfrac{2 \hat{v}^2}{243 \hat{s}^4}+\tfrac{16}{243 \hat{s}^3}+\tfrac{4}{243 \hat{s}^2}\big],
\end{align}
for $q\bar{q}$ induce process.
Here, the strong coupling constant is denoted by $g_s$.
Note, for charmonium-pair production, the processes $g+g\to J/\psi +\eta_c$ and $q+\bar{q}\to J/\psi +\eta_c$ are forbidden in the color-singlet model, due to the requirement of charge-parity $C$ conservation.
For the same reason, the amplitudes for $B_c^++B_c^{*-}$ production (Eqs. (\ref{eq_ggPV})(\ref{eq_qqPV})) vanish in the limit $m_b\to m_c$.

\section{The NLO QCD corrections}
The NLO QCD corrections to $B_c$-pair hadroproduction include virtual and real corrections, which are outlined in subsection \ref{sec_virtual} and \ref{sec_real} respectively.

\subsection{The virtual corrections}
\label{sec_virtual}
\begin{figure}
\centering
\begin{subfigure}{\textwidth}
\includegraphics[scale=0.5]{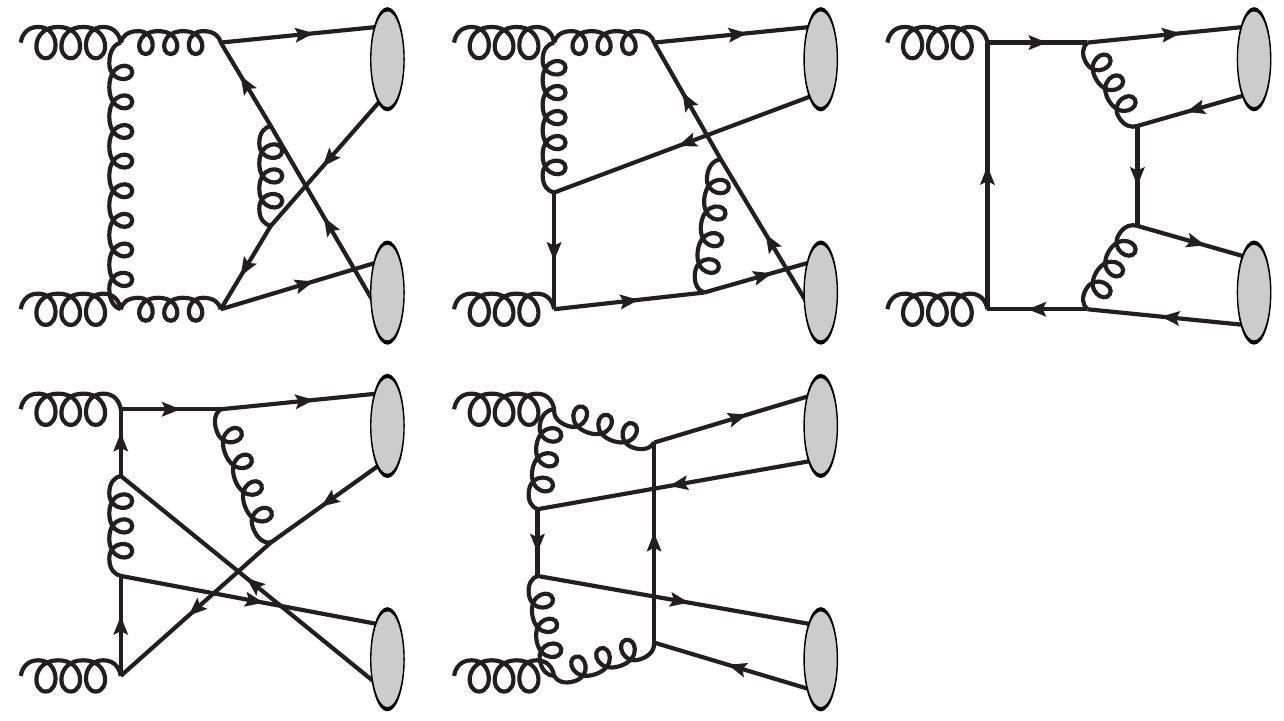}
\caption{}
\end{subfigure}
\begin{subfigure}{\textwidth}
\includegraphics[scale=0.5]{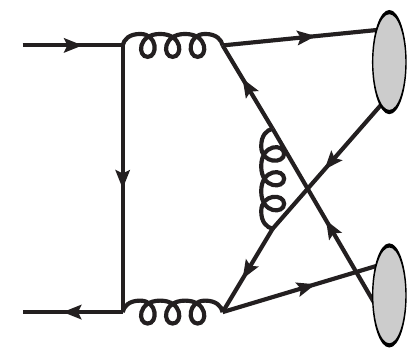}
\caption{}
\end{subfigure}
\caption{The typical hexagon diagrams for (a) $gg$-induced process, (b) $q\bar{q}$-induced process.
The remaining diagrams can be obtained by reversing the quark lines, or interchanging the initial gluons.}
\label{fig_FeynVirtual}
\end{figure}

The virtual corrections comprise about $1300$ one-loop diagrams for the $gg$ and about $100$ diagrams for the $q\bar{q}$-induced process,
the most complicated being the $38$ and $8$ hexagons for the respective channels, as shown in Fig. \ref{fig_FeynVirtual}.
The contribution of virtual corrections can be formulated as
\begin{equation}
d\hat{\sigma}_{\rm virtual}=\frac{|\Psi(0)|^4}{2s}\overline{\sum}2{\rm Re}\left(\mathcal{M}_\text{1-loop}\mathcal{M}^*_{\rm tree}\right)d{\rm PS}_2.
\end{equation}
Here, the interference term ${\rm Re}\left(\mathcal{M}_\text{1-loop}\mathcal{M}^*_{\rm tree}\right)$ contains both ultraviolet (UV) and infrared (IR) singularities.
The conventional dimensional regularization with $D=4-2\epsilon$ is employed to regularize them.
The method proposed in Refs. \cite{Kreimer:1989ke,Korner:1991sx} is used to deal with the $D$-dimensional $\gamma_5$ trace.

The UV singularities, which are contained in self-energy and trigangle diagrams, are canceled by corresponding counterterm diagrams.
The relevant renormalization constants include $Z_2$, $Z_m$, $Z_3$, $Z_l$ and $Z_g$, corresponding to heavy quark field, heavy quark mass, gluon field, light quark field, and strong coupling constant, respectively.
We define $Z_2$, $Z_m$, $Z_3$ and $Z_l$ in the on-shell (OS) scheme, $Z_g$ in the modified minimal-subtraction ($\overline{\rm MS}$) scheme.
The counter-terms are
\begin{align}
\delta Z_2^{\rm OS}=&-C_F\frac{\alpha_s}{4\pi}
\left[\frac{1}{\epsilon_{\rm UV}}+\frac{2}{\epsilon_{\rm IR}}
-3\gamma_E+3\ln\frac{4\pi\mu^2}{m_Q^2}+4\right],
\nonumber\\
 \delta Z_m^{\rm OS}=&-3C_F\frac{\alpha_s}{4\pi}
\left[\frac{1}{\epsilon_{\rm
UV}}-\gamma_E+\ln\frac{4\pi\mu^2}{m_Q^2} +\frac{4}{3}\right], \nonumber\\
 \delta Z_l^{\rm OS}=&-C_F\frac{\alpha_s}{4\pi}
\left[\frac{1}{\epsilon_{\rm UV}}-\frac{1}{\epsilon_{\rm IR}}\right], \nonumber\\
\delta Z_3^{\rm OS}=&(\beta_0^{\rm light}-2C_A)\dfrac{\alpha_s}{4\pi}
\left[ \dfrac{1}{\epsilon_{\rm UV}} -\dfrac{1}{\epsilon_{\rm IR}} \right]-\sum_{Q=b,c}\dfrac{4}{3}T_F\dfrac{\alpha_s}{4\pi}\left[\dfrac{1}{\epsilon_{\rm UV}} -\gamma_E+\ln\dfrac{4\pi\mu^2}{m^2_Q}\right],
 \nonumber\\
  \delta Z_g^{\overline{\rm MS}}=&-\frac{\beta_0}{2}\,
  \frac{\alpha_s}{4\pi}
  \left[\frac{1}{\epsilon_{\rm UV}} -\gamma_E + \ln(4\pi)
  \right].
\end{align}
Here, $\mu$ is the renormalization scale,
$\gamma_E$ is the Euler's constant,
$m_Q$ stands for $m_b$ and $m_c$ accordingly,
$\beta_0^{(\text{light})}=(11/3)C_A-(4/3)T_Fn_f^{(\text{light})}$ is the one-loop coefficiens of QCD beta function with $n_f=5$ being the number of active quarks, $n_f^\text{light}=3$ is the number of light quarks,
$C_A=3$, $C_F=4/3$ and $T_F=1/2$ are color factors.

The IR singularities in virtual corrections can be classified into two groups: the final-state-related singularities and the initial-state-related ones.
The former arise from diagrams where one final state particle exchanges a soft gluon with another on shell particle.
This type of singularities cancel each other as expected \cite{Bodwin:1994jh,Campbell:2007ws,Butenschoen:2019lef}.
The initial-state-related singularities stem from diagrams where the two initial state partons are connected by a soft gluon, or where one initial state parton is attached by a collinear gluon.
According to the Kinoshita-Lee-Nauenberg theorem \cite{Kinoshita:1962ur,Lee:1964is}, parts of initial-state-related singularities are canceled by the real corrections, and the remainings are eliminated by introducing the scale dependent PDFs.

The Coulomb singularities arise when two heavy quarks, which move with a small relative velocity, exchange a soft gluon between them.
For each $B_c$ meson, since we set $p_c=\frac{m_c}{m_b}p_b$ before the calculation of Feynman integrals, the corresponding Coulomb singularities are absent in the dimensional regularization \cite{Beneke:1997zp}.
While there is another type of Coulomb singularities: in the threshold region where $s\sim 4m_{B_c}^2$, the valence quark of $B_c^+$ moves with a small velocity relative to the valence quark of $B_c^-$. In this case, the Coulomb singularities appear as $1/\sqrt{s-4m_{B_c}^2}$ terms. We find that these terms vanish after summing up all one-loop diagrams.

\subsection{The real corrections}
\label{sec_real}
\begin{figure}
\centering
\begin{subfigure}{\textwidth}
\includegraphics[scale=0.5]{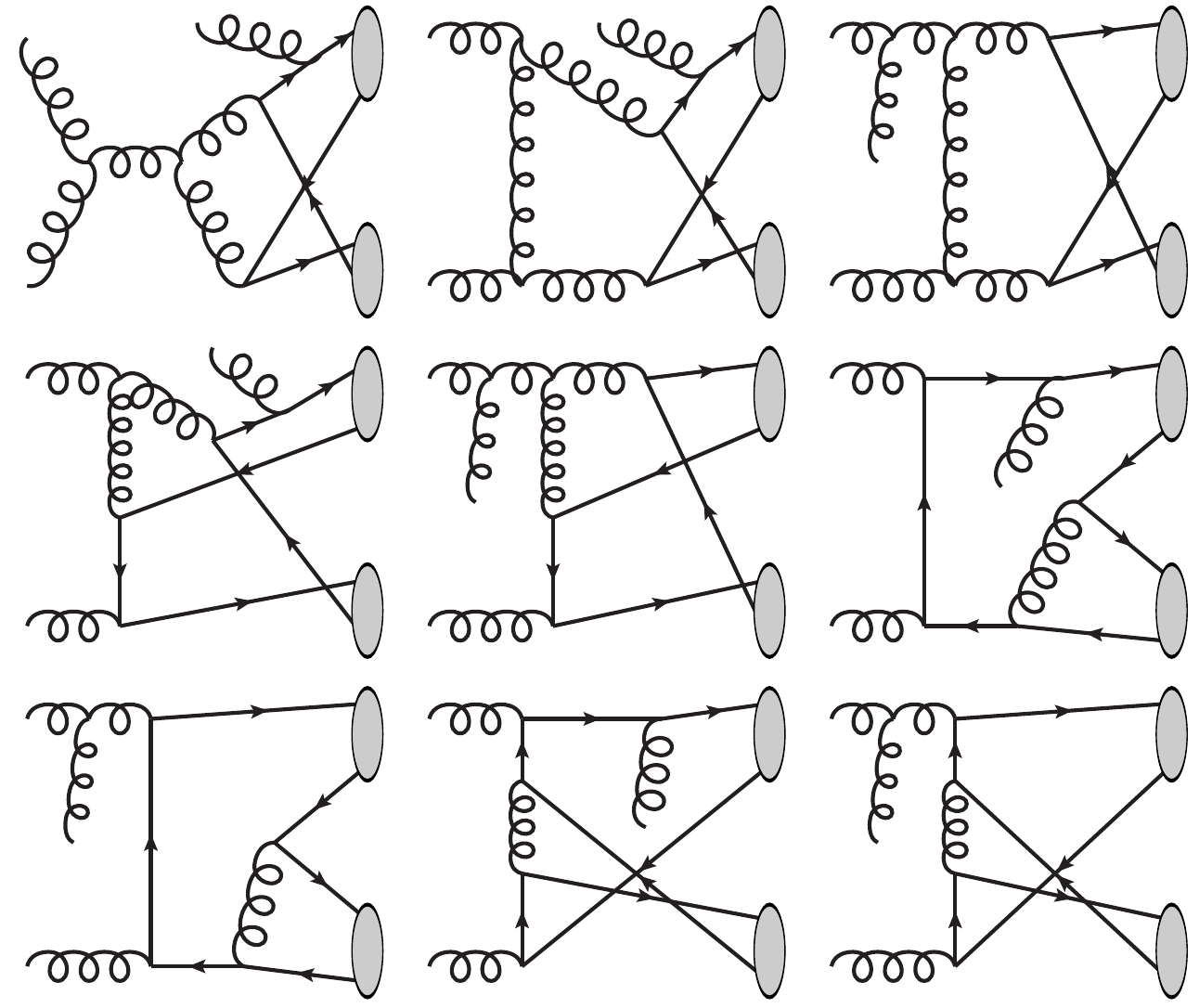}
\caption{}
\end{subfigure}
\begin{subfigure}{\textwidth}
\includegraphics[scale=0.5]{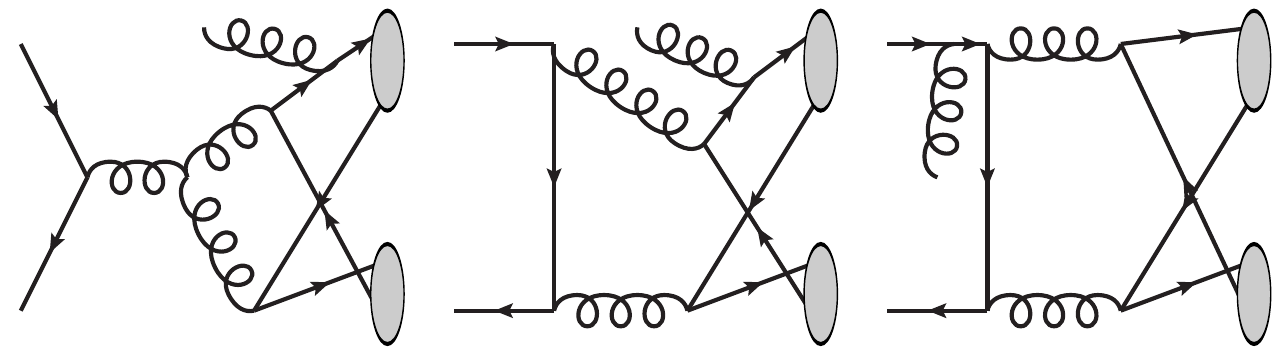}
\caption{}
\end{subfigure}
\caption{Typical Feynman diagrams for (a) $g+g\to B_c^++B_c^-+g$; (b) $q+\bar{q}\to B_c^++B_c^-+g$.
The diagrams for $g+q(\bar{q})\to B_c^++B_c^-+q(\bar{q})$ can be obtained through crossing.}
\label{fig_Feyreal}
\end{figure}

The real corrections are induced by the partonic processes
\begin{align}
&g(p_1)+g(p_2)\to B_c^+(k_1)+B_c^-(k_2)+g(p_3),\label{eq_Rprocess1} \\
&q(p_1)+\bar{q}(p_2)\to B_c^+(k_1)+B_c^-(k_2)+g(p_3),\label{eq_Rprocess2} \\
&g(p_1)+q(\bar{q})(p_2)\to B_c^+(k_1)+B_c^-(k_2)+q(\bar{q})(p_3).\label{eq_Rprocess3}
\end{align}
Here, processes (\ref{eq_Rprocess1})(\ref{eq_Rprocess2}) are the gluon bremsstrahlung corrections to corresponding LO processes.
Process (\ref{eq_Rprocess3}) indicates a new production channel, namely the gluon-quark scattering.
The typical Feynman diagrams for processes (\ref{eq_Rprocess1})(\ref{eq_Rprocess2}) are shown in Fig. \ref{fig_Feyreal}.
The diagrams for process (\ref{eq_Rprocess3}) can be obtained through crossing.

In the calculation of real corrections, the Catani–Seymour (CS) dipole subtraction method \cite{Catani:1996vz,Catani:2002hc} is adopted to handle the IR singularities.
The general idea of this method is to rewrite the NLO corrections as
\begin{equation}
\hat{\sigma}_{\rm NLO}=\int_\text{3-body}\left[d\hat{\sigma}_{\rm real}-d\hat{\sigma}_A\right]+\int_\text{2-body} \left[d\hat{\sigma}_{\rm virtual}+d\hat{\sigma}_C+\int_\text{1-body} d\hat{\sigma}_A\right].
\label{eq_Rsubf}
\end{equation}
Here, $d\hat{\sigma}_A$ is an auxiliary dipole subtraction term which possesses the same pointwise singular behavior as $d\hat{\sigma}_{\rm real}$,
$d\hat{\sigma}_C$ is the collinear subtraction term which originates from the renormalization of PDFs.
As $d\hat{\sigma}_A$ acts as a local counter term for $d\hat{\sigma}_{\rm real}$, the first integral of Eq. (\ref{eq_Rsubf}) is nonsingular at every points of phase space, and can be evaluated numerically in four dimensions.
On the other hand, the integration of $d\hat{\sigma}_A$ over one-body subspace, i.e. the $\int_\text{1-body} d\hat{\sigma}_A$, can be carried out analytically in $D$ dimensions.
As a result, and the IR singularities in real corrections appear as pole terms of $1/\epsilon^n$.
After adding up $\int_\text{1-body} d\hat{\sigma}_A$, $d\hat{\sigma}_{\rm virtual}$ and $d\hat{\sigma}_C$, all IR singularities are eliminated as expected.

The dipole subtraction term $d\hat{\sigma}_A$ can be constructed term by term according to all possible emitter-spectator pairs.
While in our case, the contributions of final-state-emission diagrams cancel each other \cite{Bodwin:1994jh,Campbell:2007ws,Butenschoen:2019lef}.
Hence, only the dipoles where both emitter and spectator are in the initial state need to be taken into account.
Following the implementation of Ref. \cite{Catani:1996vz}, the $d\hat{\sigma}_A$ is constructed as
\begin{align}
d\hat{\sigma}_A^{1+2\to [b\bar{c}]+[\bar{b}c]+3}=&\frac{1}{2p_1\cdot p_2}\bigg\{\frac{g_s^2}{p_1\cdot p_3}\frac{P_{11^\prime}(x)}{x}\overline{\sum}\left|\mathcal{M}_{\rm tree}^{1^\prime +2\to [b\bar{c}]+[\bar{b}c]}(\tilde{p}_1,\tilde{p}_2;\tilde{k}_1,\tilde{k}_2)\right|^2\nonumber \\
+&\frac{g_s^2}{p_2\cdot p_3}\frac{P_{22^\prime}(x)}{x}\overline{\sum}\left|\mathcal{M}_{\rm tree}^{1+2^\prime\to [b\bar{c}]+[\bar{b}c]}(\tilde{p}_1,\tilde{p}_2;\tilde{k}_1,\tilde{k}_2)\right|^2\bigg\}d{\rm PS}_3,
\label{eq_subterm}
\end{align}
where $x=1-(p_1\cdot p_3+p_2\cdot p_3)/(p_1\cdot p_2)$ is the momentum fraction at the collinear limit,
$P_{ab}(x)$ is the spin average of the Altarelli-Parisi splitting function, which are of the form
\begin{align}
&P_{qq}(x)=C_F\tfrac{1+x^2}{1-x},\\
&P_{qg}(x)=C_F\tfrac{1+(1-x)^2}{x},\\
&P_{gq}(x)=T_F[1-2x(1-x)],\\
&P_{gg}(x)=2C_A[\tfrac{x}{1-x}+\tfrac{1-x}{x}+x(1-x) ].
\end{align}
The CS projection, which projects the three-body phase space into the two-body one, is implemented as
\begin{align}
&\text{emitter}: \tilde{p}_{1,2}=x p_{1,2},\\
&\text{spectator}: \tilde{p}_{2,1}=p_{2,1},\\
&\text{others}: \tilde{k}_{1,2}=k_{1,2}-\tfrac{2k_{1,2}\cdot(K+\tilde{K})}{(K+\tilde{K})^2}(K+\tilde{K})+\tfrac{2k_{1,2}\cdot K}{K^2}\tilde{K},
\end{align}
with $K=p_1+p_2-p_3$, $\tilde{K}=\tilde{p}_1+\tilde{p}_2$.

For the second term of Eq. (\ref{eq_Rsubf}), we have
\begin{align}
&d\hat{\sigma}_C^{1+2\to [b\bar{c}]+[\bar{b}c]+3}+\int_1 d\hat{\sigma}_A^{1+2\to [b\bar{c}]+[\bar{b}c]+3}
=\int^1_0 dx\ \frac{\alpha_s}{2\pi}\frac{1}{\Gamma(1-\epsilon)}\left(\frac{4\pi \mu^2}{2p_1\cdot p_2}\right)^\epsilon\nonumber \\
&
\times\bigg[ d\hat{\sigma}_{\rm LO}^{1^\prime+2\to [b\bar{c}]+[\bar{b}c]}V_{11^\prime}(x,p_1,p_2,\mu_F)+d\hat{\sigma}_{\rm LO}^{1+2^\prime\to [b\bar{c}]+[\bar{b}c]}V_{22^\prime}(x,p_1,p_2,\mu_F)\bigg],
\end{align}
and according to Ref. \cite{Catani:1996vz}
\begin{align}
V_{qq}(x,p_1,p_2,\mu_F)=&\tfrac{4}{3}(1-x)-\tfrac{8}{3}(1+x)\ln(1-x)+\tfrac{8}{3}\ln\tfrac{\sqrt{2p_1\cdot p_2}}{\mu_F}[\tfrac{1+x^2}{1-x}]_+\nonumber \\
+&\tfrac{16}{3}[\tfrac{\ln(1-x)}{1-x}]_++(\tfrac{4}{3}\tfrac{1}{\epsilon^2}+\tfrac{2}{\epsilon}-\tfrac{2}{9}\pi^2)\delta(1-x),\\
V_{gq}(x,p_1,p_2,\mu_F)=&[x^2+(1-x)^2][\ln(1-x)+\ln\tfrac{\sqrt{2p_1\cdot p_2}}{\mu_F}]+x(1-x),\\
V_{qg}(x,p_1,p_2,\mu_F)=&\tfrac{8}{3}\tfrac{1+(1-x)^2}{x}[\ln(1-x)+\ln\tfrac{\sqrt{2p_1\cdot p_2}}{\mu_F}]+\tfrac{4}{3}x,\\
V_{gg}(x,p_1,p_2,\mu_F)=&12[-1+x(1-x)+\tfrac{1-x}{x}][\ln(1-x)+\ln\tfrac{\sqrt{2p_1\cdot p_2}}{\mu_F}]\nonumber \\
+&12[\tfrac{\ln(1-x)}{1-x}]_++12[\tfrac{1}{1-x}]_+\ln\tfrac{\sqrt{2p_1\cdot p_2}}{\mu_F}\nonumber \\
+&[\tfrac{3}{\epsilon^2}+(\tfrac{11}{2}-\tfrac{n_f^\text{light}}{3})(\tfrac{1}{\epsilon}+2\ln\tfrac{\sqrt{2p_1\cdot p_2}}{\mu_F})-\tfrac{\pi^2}{2}]\delta(1-x).
\end{align}
Here, $\mu_F$ is the factorization scale, the `$+$'-distribution is defined as
\begin{equation}
\int_0^1dx\ g(x)[f(x)]_+=\int_0^1dx\ [g(x)-g(1)]f(x).
\end{equation}

\section{Numerical results}
\subsection{Input parameters}
We consider $B_c$-pair production at the LHC with $\sqrt{s}=14$ TeV.
The hadron-level cross sections can be obtained by convoluting the partonic cross sections with the proton PDFs
\begin{align}
d&\sigma^{p+p\to B_c^++B_c^-+X}=\int dx_1dx_2\ f_g(x_1) f_g(x_2) d\hat{\sigma}^{g+g\to B_c^++B_c^-+(g)}\nonumber \\
&+\sum_{q=u,d,s}\int dx_1dx_2\ \left[f_q(x_1) f_{\bar{q}}(x_2) d\hat{\sigma}^{q+\bar{q}\to B_c^++B_c^-+(g)}+\ f_{\bar{q}}(x_1) f_q(x_2) d\hat{\sigma}^{\bar{q}+q\to B_c^++B_c^-+(g)}\right]\nonumber \\
&+\sum_{\substack{q=u,d,s,\\ \ \ \bar{u},\bar{d},\bar{s}}}\int dx_1dx_2\ \left[f_g(x_1) f_{q}(x_2) d\hat{\sigma}^{q+g\to B_c^++B_c^-+q}+f_q(x_1) f_g(x_2) d\hat{\sigma}^{g+q\to B_c^++B_c^-+q}\right].
\label{eq_ppcro}
\end{align}
Here, all partonic cross sections are evaluated at $\mathcal{O}(\alpha_s^5)$ accuracy.
In our numerical studies, the CT18 NLO parametrization with $\alpha_s(M_Z)=0.118$ \cite{Hou:2019efy} is used for the evaluation of PDFs and $\alpha_s$.
The renormalization and factorization scales are chosen as $\mu_0\le\mu_R=\mu_F\le 2\mu_0$, where the central scale $\mu_0$ is taken to be the half of the summed transverse masses of the two $B_c$ mesons, i.e. $\mu_0=\left(\sqrt{m_{B_c}^2+p_{\text{T},1}^2}+\sqrt{m_{B_c}^2+p_{\text{T},2}^2}\right)/2$.

Other parameters in numerical evaluation go as follows
\begin{equation}
m_b=4.60\ \text{GeV},\quad m_c=1.49\ \text{GeV},\quad |\Psi(0)|^2=0.174\ \text{GeV}^3.
\end{equation}
Here, the pole masses of heavy quark are obtained through
\begin{equation}
m_Q=\bar{m}_Q\left(1+C_F\frac{\alpha_s(\bar{m}_Q)}{\pi}\right),
\end{equation}
with the $\overline{\text{MS}}$ masses $\bar{m}_b=4.18\ \text{GeV}$ and $\bar{m}_c=1.27\ \text{GeV}$ \cite{ParticleDataGroup:2020ssz} as input.
The $B_c$ wave function at the origin is is estimated from the ${}^1S_0-{}^3S_1$ splitting \cite{Eichten:1994gt}
\begin{equation}
|\Psi(0)|^2=\frac{9m_bm_c}{21\pi \alpha_s}(m_{B_c^*}-m_{B_c}),
\end{equation}
with the lattice calculation result $m_{B_c^*}-m_{B_c}= 53$ MeV \cite{Gregory:2010gm} as input.

Due to the finite coverage of detectors, proper kinematic cuts should be imposed on the final state particles.
Here we employ two typical sets of cuts, which corresponding to the LHCb \cite{LHCb:2014mvo} and ATLAS \cite{ATLAS:2019jpi} acceptance of $B_c$ meson respectively:
\begin{itemize}
\item LHCb cuts: $1\le p_{\text{T},1},p_{\text{T},2} \le 20$ GeV, $2.0\le y_1,y_2\le 4.5$,
\item ATLAS cuts: $ p_{\text{T},1},p_{\text{T},2} \ge 13$ GeV, $-2.3\le y_1,y_2 \le 2.3$,
\end{itemize}
where $p_{\text{T},i}$ and $y_i$ with $i=1,2$ denote the transverse momentum and rapidity of each $B_c$ meson respectively.
Note, according to Ref. \cite{CMS:2020rcj}, the CMS acceptance cuts of $B_c$ meson are similar to the ATLAS ones.

\subsection{Integrated cross sections}
The LO and NLO cross sections for $B_c^++B_c^-$, $B_c^++B_c^{*-}$ and $B_c^{*+}+B_c^{*-}$ production with different cuts are shown in Table \ref{tab_intCro}, wherein,  the central values refer to the results at $\mu_{R/F}=\mu_0$, the super- and subscripts corresponding to the results at $\mu_{R/F}=\mu_0/2$ and $\mu_{R/F}=2\mu_0$ respectively.
Besides the total cross sections, the partial cross sections for $gg$-, $q\bar{q}$-, and $gq(\bar{q})$-induced processes,
i.e. the  first, second, and third lines of Eq. (\ref{eq_ppcro}),
are presented as well.
About Table \ref{tab_intCro}, there are some points remarkable which are as follows:\\
\indent  (i) Due to the high gluon flux at low $x$, the $B_c$-pair production at the LHC is dominated by the $gg$-induced processes as expected.
While in some cases, the contributions of other partonic processes are also non-negligible.
For example, for $B_c^{*+}+B_c^{*-}$ production with ATLAS cuts, the contribution of $gq(\bar{q})$-induced process reaches $20\%$ level at the NLO.\\
\indent (ii) Similar to the situation of $J/\psi$-pair production \cite{Sun:2014gca},  the NLO corrections here may dominate over the LO contributions,
and the theoretical uncertainties induced by energy scale is even larger at NLO than that at the LO.
We propose a potential explanation based on the kinematic analysis.
We find that the bulk of the cross sections arise from the regime where the invariant mass of the $B_c$-pair $M_{2B_c}$ is small.
At LO, the two body kinematic forces the $B_c$ mesons to be back to back in the transverse momentum plane,
hence the $p_{\rm T}$ cut on each $B_c$ meson lead to a constraint of $M_{2B_c}\ge 2\sqrt{m_{B_c}^2+p_{\text{T,min}}^2}$.
While at NLO, the three body real emission processes are involved.
In this case, the $B_c$-pair can be produced near the mass threshold, i.e. $M_{2B_c}\sim 2m_{B_c}$, as long as the recoil parton possesses large enough transverse momentum.
The threshold production effect appearing at the real corrections accounts for the extraordinary large NLO corrections.\\
\indent (iii) There are negative cross section problems in $q\bar{q}$- and $gq(\bar{q})$- induced processes.
For the former, the partial cross sections become negative when $\mu_{R,F}$ approaching $\mu_0/2$,
and the large negative corrections can be traced back to the virtual corrections.
For the latter, the partial cross sections become negative when $\mu_{R,F}$ approaching $2\mu_0$,
and the large negative corrections can be traced back to the over-subtraction of collinear singularities inside the PDFs \cite{ColpaniSerri:2021bla}.
Since the contributions of these two processes are not dominant, the total ($pp$) cross sections keep positive.
Hence, in our calculations, no additional treatment is applied to cure the negative cross section problems.

\begin{table}
   \centering
  \caption{The LO and NLO integrated cross sections for $B_c$-pair production at the LHC. The LO results agree with those in Ref. \cite{Li:2009ug} after taking the same inputs.}
  \label{tab_intCro}
    \begin{tabular}{|p{2cm}<{\centering}|p{2cm}<{\centering}|p{3cm}<{\centering}|p{3cm}<{\centering}|p{3cm}<{\centering}|p{3cm}<{\centering}|}
    \hline
     \multicolumn{2}{|c|}{channel}&
    \multicolumn{2}{c|}{LHCb cuts (in nb)}&\multicolumn{2}{c|}{ATLAS cuts (in pb)}
    \cr\hline
  final state&initial state &LO&NLO&LO&NLO\cr
    \hline
   \multirow{4}{*}{$B_c^++B_c^-$}& $gg$ & $2.39^{+0.07}_{-0.36}\times 10^{-1}$ & $5.68^{+1.27}_{-0.83}\times 10^{-1}$ & $4.09^{+1.04}_{-0.92}\times 10^{-1}$ & $2.99^{+1.89}_{-1.04}$\cr\cline{2-6}
  & $q\bar{q}$ & $1.90^{+0.69}_{-0.44}\times 10^{-4}$ & $1.75^{-1.55}_{+0.48}\times 10^{-4}$ & $4.66^{+1.55}_{-1.11}\times 10^{-3}$ & $2.93^{-3.67}_{+1.32}\times 10^{-3}$\cr\cline{2-6}
 &  $gq(\bar{q})$ & --- & $0.16^{+7.75}_{-2.83}\times 10^{-2}$ & --- & $6.99^{+6.10}_{-3.12}\times 10^{-1}$\cr\cline{2-6}
 &  total ($pp$)  & $2.39^{+0.07}_{-0.36}\times 10^{-1}$ & $5.70^{+2.05}_{-1.11}\times10^{-1}$ & $4.14^{+1.05}_{-0.93}\times 10^{-1}$ & $3.69^{+2.50}_{-1.35}$\cr \hline \hline
     \multirow{4}{*}{$B_c^++B_c^{*-}$}& $gg$ & $3.43^{+0.27}_{-0.58}\times 10^{-2}$ & $8.34^{+2.80}_{-1.56}\times 10^{-2}$ & $3.46^{+0.89}_{-0.79}\times 10^{-1}$ & $3.20^{-0.09}_{+0.05}\times 10^{-1}$\cr\cline{2-6}
&   $q\bar{q}$ & $5.36^{+1.94}_{-1.18}\times 10^{-3}$ & $3.22^{-5.92}_{+2.12}\times 10^{-3}$ & $2.12^{+0.69}_{-0.50}\times 10^{-2}$ & $1.36^{-1.67}_{+0.61}\times 10^{-2}$\cr\cline{2-6}
 &  $gq(\bar{q})$ & --- & $0.68^{+1.83}_{-0.76}\times 10^{-2}$ &  --- & $1.16^{+9.56}_{-3.63}\times10^{-2}$\cr\cline{2-6}
&   total ($pp$) & $3.96^{+0.46}_{-0.70}\times 10^{-2}$ & $9.34^{+4.04}_{-2.12}\times 10^{-2}$ & $3.67^{+0.96}_{-0.84}\times10^{-1}$ & $3.45^{+0.70}_{-0.25}\times10^{-1}$\cr
    \hline \hline
   \multirow{4}{*}{$B_c^{*+}+B_c^{*-}$}&   $gg$ & $4.06^{+0.23}_{-0.65}\times 10^{-1}$ & $1.10^{+0.47}_{-0.24}$ & $3.00^{+0.78}_{-0.68}$ & $10.2^{+5.9}_{-3.1}$\cr\cline{2-6}
&   $q\bar{q}$ & $4.17^{+1.51}_{-0.92}\times 10^{-2}$ & $1.90^{-5.01}_{+1.85}\times 10^{-2}$ & $1.76^{+0.57}_{-0.42}\times 10^{-1}$ & $0.66^{-1.65}_{+0.66}\times 10^{-1}$\cr\cline{2-6}
&   $gq(\bar{q})$ & --- & $1.56^{+2.62}_{-1.16}\times 10^{-1}$ & --- & $2.62^{+2.70}_{-1.32}$\cr\cline{2-6}
&   total ($pp$) & $4.48^{+0.38}_{-0.74}\times 10^{-1}$ & $1.28^{+0.68}_{-0.34}$ & $3.17^{+0.84}_{-0.72}$ & $12.8^{+8.4}_{-4.3}$\cr
    \hline
    \end{tabular}
\end{table}

Since $B_c^*$ almost all decays to $B_c$, a prediction on $B_c$-pair candidates should sum over all possible final states.
By introducing the final-state-summed cross section as $\sigma^{\rm FSS}=\sigma^{B_c^++B_c^-}+2\sigma^{B_c^++B_c^{*-}}+\sigma^{B_c^{*+}+B_c^{*-}}$,
we have
\begin{align}
&\sigma^{\rm FSS}_{\rm LO}=7.67^{+0.55}_{-1.24}\times 10^{-1}\; \text{nb},\ \sigma^{\rm FSS}_{\rm NLO}=2.03^{+0.97}_{-0.49}\; \text{nb},\ \text{for LHCb cuts}; \\
&\sigma^{\rm FSS}_{\rm LO}=4.32^{+1.14}_{-0.99}\times 10^{-3}\; \text{nb},\ \sigma^{\rm FSS}_{\rm NLO}=1.72^{+1.11}_{-0.57}\times 10^{-2}\; \text{nb},\ \text{for ATLAS cuts}.
\end{align}
The High Luminosity LHC (HL-LHC) will substantially increase the amount of proton-proton collisions delivered to the LHC experiments, with a planned integrated luminosity of $4000\; \text{fb}^{-1}$ for ATLAS experiment \cite{SanchezCruz:2022whm}, and $300\; \text{fb}^{-1}$ for LHCb \cite{Kholodenko:2022anr}.
Assuming $B_c$ is reconstructed through the decay $B_c^\pm\to J/\psi\pi^\pm$, whose branching fraction is predicted to be $0.5\%$ \cite{Chang:1992pt}, and $J/\psi$ is reconstructed through $J/\psi\to l^+l^-(l=e,\mu)$ with a branching fraction of about $12\%$ \cite{ParticleDataGroup:2020ssz}, the number of the reconstructed $B_c$-pair candidates is about $166$ -- $324$ for LHCb experiment, and about $16$ -- $40$ for ATLAS experiment.

\subsection{Differential cross sections}
\begin{figure}
\centering
\begin{subfigure}{0.32\textwidth}
\includegraphics[width=\textwidth]{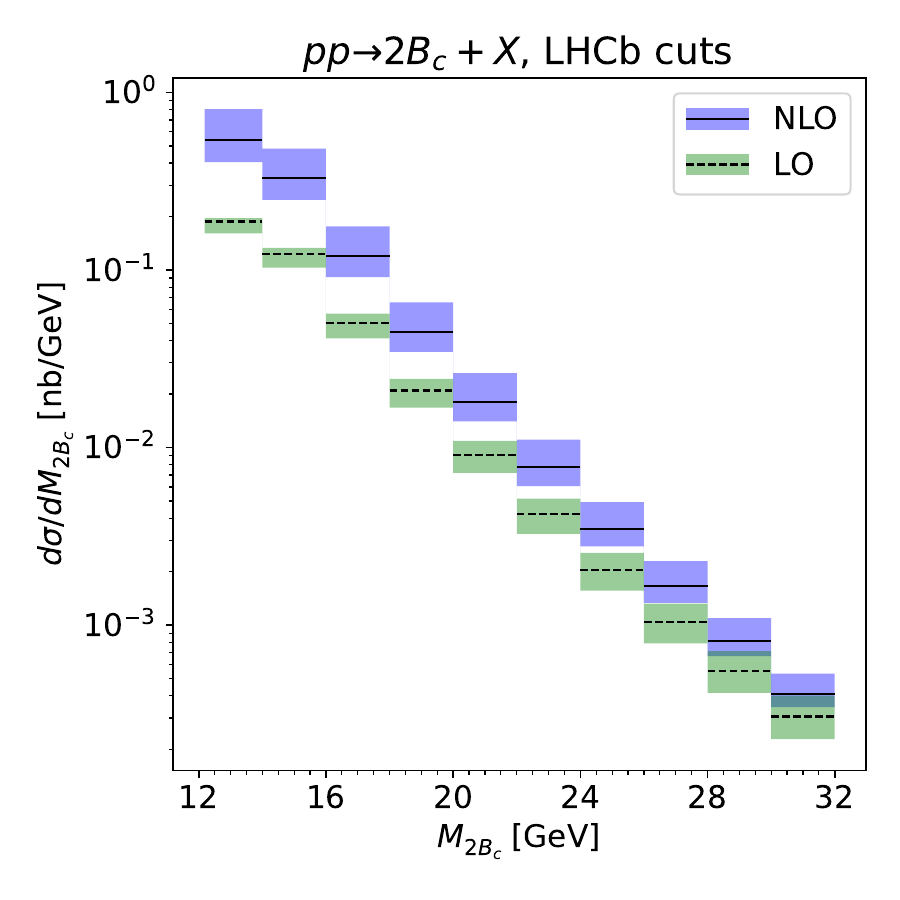}
\caption{}
\end{subfigure}
\begin{subfigure}{0.32\textwidth}
\includegraphics[width=\textwidth]{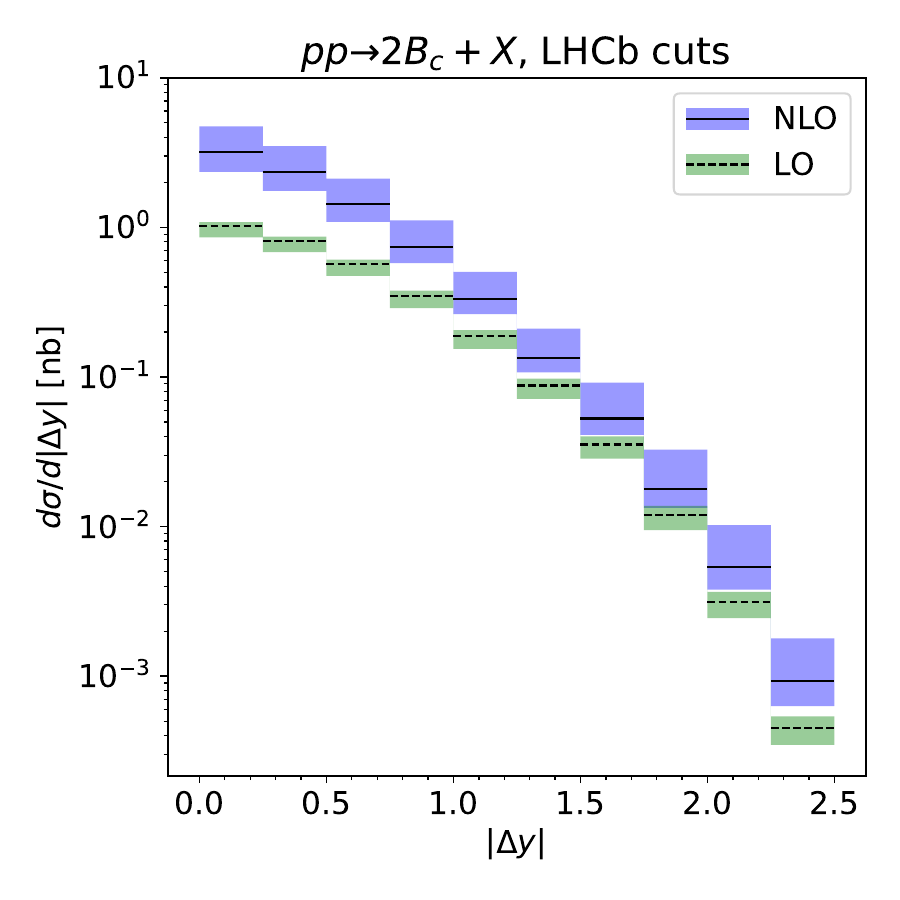}
\caption{}
\end{subfigure}
\begin{subfigure}{0.32\textwidth}
\includegraphics[width=\textwidth]{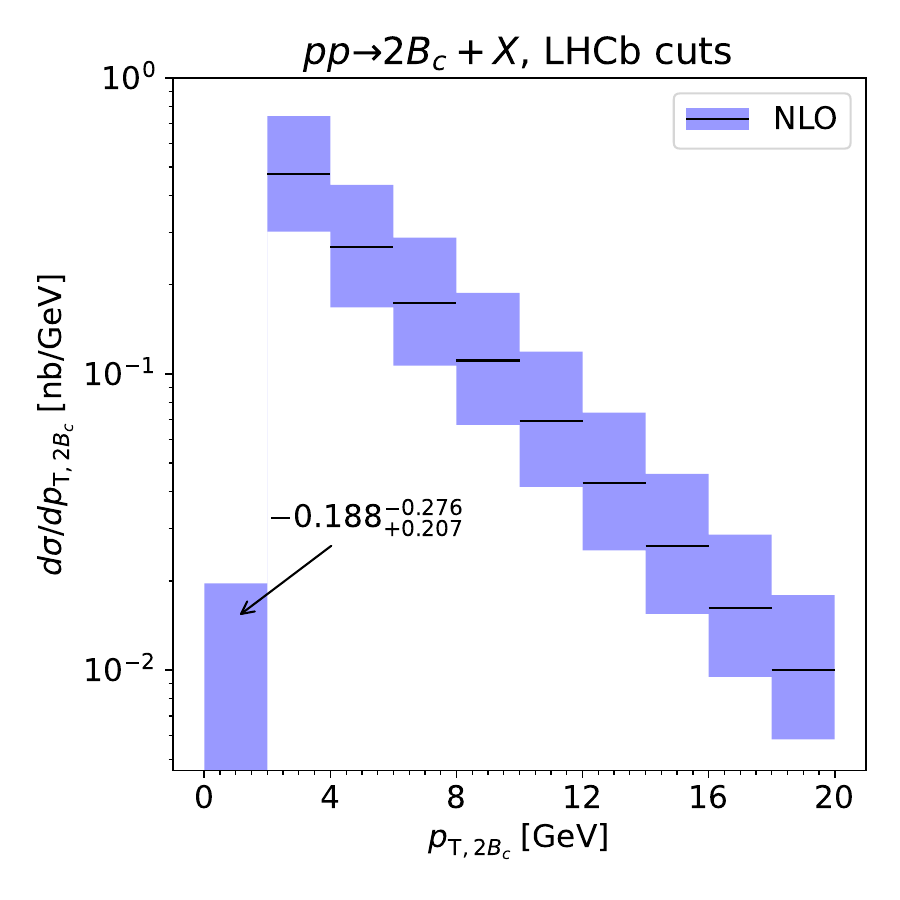}
\caption{}
\end{subfigure}
\caption{The differential distributions in (a) the invariant mass of the $B_c$-pair $M_{2B_c}$, (b) the absolute value of rapidity difference of the $B_c$-pair $|\Delta y|$, (c) the transverse momentum of the $B_c$-pair $p_{\text{T},2B_c}$.}
\label{fig_3dis}
\end{figure}

As the number of events corresponding to LHCb experiment is promising, it is worthy to perform a more elaborate phenomenological analysis.
The differential distributions in various variables with LHCb cuts are shown in Fig. \ref{fig_3dis},
wherein, the LO and NLO predictions at $\mu_{R/F}=\mu_0$ are represented by dashed and solid lines respectively; the theoretical uncertainties, which are estimated by varying $\mu_{R/F}$ in the range $[\mu_0/2,2\mu_0]$,  are represented by color bands.

The distributions in $M_{2B_c}$, the invariant mass of the $B_c$-pair, are shown in Fig. \ref{fig_3dis}(a).
It can be seen that as $M_{2B_c}$ increases from about $12\; \text{GeV}$ to $32\; \text{GeV}$, both LO and NLO differential cross sections drop steadily.
Although the NLO corrections are positive everywhere in the plotted $M_{2B_c}$ range, the relative size of NLO corrections over LO contributions decreases with increasing $M_{2B_c}$.

The distributions in $|\Delta y|$, the absolute value of rapidity difference of the two $B_c$ mesons, i.e. $|\Delta y|=|y_1-y_2|$, are shown in Fig. \ref{fig_3dis}(b).
Similar to the $M_{2B_c}$ distributions, the differential cross sections drop steadily with increasing $|\Delta y|$, and the NLO corrections are more significant in the small $|\Delta y|$ region.

The distributions in $p_{\text{T},2B_c}$, the transverse momentum of the $B_c$-pair, are shown in Fig. \ref{fig_3dis}(c).
Since the $B_c$-pair with non-vanishing $p_{\text{T},2B_c}$ can only be produced in three-body processes, the second to the last bins only receive contribution from the $d\hat{\sigma}_{\rm real}$ term of Eq. (\ref{eq_Rsubf}) .
For the first bin, the first term of Eq. (\ref{eq_Rsubf}) is negative, and larger in absolute value than the second term of Eq. (\ref{eq_Rsubf}).
Hence the cross sections in the first bin can be negative at some energy scale.
A resummation of the logarithms of $p_{\text{T},2B_c}^2/s$ may help to solve this problem. This asks for a further investigation.

\section{Summary}
In this work, we investigate the $B_c$-pair production in proton-proton collisions at the NLO accuracy in the framework of NRQCD factorization formalism.
Various of $S$-wave $B_c$ states, including configurations of $B_c^++B_c^-$, $B_c^++B_c^{*-}$ and $B_c^{*+}+B_c^{*-}$, are taken into account.
The total cross sections and the differential cross sections versus various variables at the LHC with $\sqrt{s}=14\; \text{TeV}$ are evaluated and presented in Table \ref{tab_intCro} and Figure \ref{fig_3dis}.

Numerical results show that, after including the NLO corrections, the total cross sections are significantly enhanced, and their dependences on renormalization and factorization scales are increased as well.
We propose a potential explanation for the poor perturbative behavior, while further investigation is still needed.
We also discuss the negative cross section problems encountered in the calculation.

Since $B_c^*$ almost always decays to $B_c$, a prediction on $B_c$-pair candidates should sum over the  $B_c^++B_c^-$, $B_c^++B_c^{*-}$, $B_c^{*+}+B_c^-$ and $B_c^{*+}+B_c^{*-}$ production rates.
As a result, we obtain $\sigma^{\rm FSS}_{\rm NLO}=2.03^{+0.97}_{-0.49}\; \text{nb}$ for LHCb experiment, and $\sigma^{\rm FSS}_{\rm NLO}=1.72^{+1.11}_{-0.57}\times 10^{-2}\; \text{nb}$ for ATLAS experiment.
Assuming $B_c$ is reconstructed through $B_c^\pm\to J/\psi\pi^\pm$, and $J/\psi$ is reconstructed through $J/\psi\to l^+l^-(l=e,\mu)$, the reconstructed $B_c$-pair candidates under the HL-LHC luminosity may reach $166$ -- $324$ for LHCb experiment, and $16$ -- $40$ for ATLAS experiment.

\vspace{2.0cm} {\bf Acknowledgments}

This work was supported in part by the National Key Research and Development Program of China under Contracts No. 2020YFA0406400, by the National Natural Science Foundation of China(NSFC) under the Grants 11975236, 12205061 and 12235008.

\vspace{2.0cm}

\end{document}